\input epsf

\documentstyle[12pt]{article}
\begin{document}
\title{Lepton Flavour Violating decays of Higgs bosons
beyond the Standard Model}
\author{ J. Lorenzo Diaz-Cruz$^1$ and J.J. Toscano$^2$  \\
  1) Instituto de Fisica, BUAP,  \\
 Ap. Postal J-48, 72500 Puebla, Pue., Mexico \\
  2) Facultad de Ciencias Fisico-Matematicas, BUAP, \\
Ap. Postal 1152, Puebla, Pue., Mexico \\
      }
\maketitle
\begin{abstract}
We evaluate the Lepton Flavor Violating (LFV) decays of Higgs bosons
$H\to l^+_il^-_j$ in several extensions of the SM, including both the 
effective lagrangian approach and several specific models.
For the effective lagrangian case, we focus on the dimension-6 
operators that induce LFV vertices for the Higgs and/or Z bosons. 
For those operators whose coefficients can not be constrained by
present data, the LFV Higgs decays could have
a Branching Ratio (B.R.) of order $10^{-1}-10^{-2}$, which can be detected at 
future hadron colliders. Those cases where 
current bounds on LFV transition apply, induce
strong bounds for $e-\mu$ transitions, and the resulting B.R. for
the decay $H\to e\mu$ is of order $10^{-9}$; however, even in this case
the bounds involving the tau are considerably weaker,
and allow the modes 
$H\to \tau \mu / \tau e$ to have a B.R. of the order $10^{-1}$.
In the case of the general Two-Higgs doublet model, we also obtain
a B.R. of the order $10^{-1}$ for the modes $H\to \tau \mu / \tau e$,
whereas for the SM with massive neutrinos
and the Minimal SUSY-SM, the B.R.'s of the corresponding LFV Higgs decays  
are very suppressed.

\end{abstract}





\section {Introduction.}
The separate conservation of the lepton number 
($L_i= L_e, L_{\mu}, L_{\tau}$) can be
considered one of the central features of the Standard Model (SM)
of electroweak interactions. This result follows automatically
from the asignment of quantum numbers to the fermions
contained in the model.
However, it is also known that lepton number could be violated in
many of its extensions; for instance by just providing neutrinos
with a mass, either through dimension-5 operators or
Higgs triplets, can lead to violation of lepton number, which can manifest
itself through several signatures. Although the
particle data book of 1998  \cite{partdat} does not report the presence of 
any new physics beyond the SM, the recent results presented  by the
Super-Kamiokande collaboration \cite{kamiok},
showing evidence for neutrino oscillations, indicate that 
lepton number is not conserved in nature and thus the lepton 
sector of the SM requires some modification to account for the
pattern of neutrino mixing that the data suggest.

 Once the Kamiokande data is confirmed as evidence
for Lepton Flavour Violation (LFV),
it is interesting to find other processes where 
such violation can be present.
Current experimental bounds on LFV transitions severely constraint
most hypothetical sources of lepton flavour non-conservation \cite{partdat}.
However, since the Higgs boson is the only part of the SM that 
has not been detected \cite{hhunt}, it is pertinent to ask if there
could be a connection or mixing
between the Higgs sector and the mechanism
responsible for the non-conservation of lepton number, and to find out wether
some remmant effect could show up in Higgs decays, which may
be detectable at present or future colliders.

In this paper, we are interested in studying the decays of the Higgs boson 
$H\to l^+_i l^-_j$, as a possible signal of LFV.
We shall work first in a model-independent approach,
using the Effective lagrangian extension of the SM.
We shall focus on the dimension-six operators that induce LFV
interactions of the Z and Higgs bosons.
Then, we also consider several specific models
where LFV Higgs decays may arise; namely,
the Standard Model with massive 
neutrinos, the general Two-Higgs doublet model (THDM-III),
and the Minimal SUSY extension of the SM (MSSM).

The main result of this work is that the decays $H\to \tau e / \tau \mu$
are generally permited to have B.R.'s of order 0.1, for both the effective 
lagrangian and THDM-III cases,
whereas the mode $H\to e\mu$ can reach at most a B.R. of order $10^{-3}$, 
only for some specific cases; however both scenarios can be
tested at future
hadron colliders. On the other hand, we find that
LFV Higgs decays are highly suppressed
for the SM with massive neutrinos and the MSSM.

\section { Model-independent results.}
The effective lagrangian approach has been used as a mean to
describe the effects of new physics in a model-independent fashion.
These effects are parametrized through effective
operators of dimension higher than four \cite{leffrev}.
 The effective Lagrangian, up to dimension-six,
is conveniently written as 
\begin{equation}
\label{eq1}
{\cal L}_{eff}={\cal L}_{SM}+\sum_{nij}\frac{\alpha^{ij}_n}{\Lambda^2}O^{ij}_n,
\end{equation}
where $i,j$, ($=1,2,3$) denote flavour indices, $n$ runs over the number of independent
operators. The scale $\Lambda$ is associated with the onset of new physics.
The coefficients $\alpha^{ij}_n$ are undetermined, and should in principle 
be calculable from the fundamental theory, but for phenomenologycal studies,
it suffices to use the bounds obtained from present experimental data. 

The conservation of lepton number, which is automatic in the SM,
is lost in general with the inclusion of higher-dimensional operators.
For instance, the dimension-5 operator: 
$O^5_{ij}= \bar{\tilde{L_i}} \tau_a L_j \tilde{\Phi} \tau_a \Phi $,
where  $L_i$ denotes the Lepton doublets and $\Phi$ 
is the Higgs doublet, can generate neutrino masses of Majorana-type
which violate lepton number by two units;
 however, one can verify that this operator
does not generate LFV interaction of the neutral Higgs boson.

Most studies of LFV in the effective lagrangian context, have focused on 
establishing bounds on the scale $\Lambda$, assuming for the coefficients
the value $\alpha^{ij}_n=1$.  Although it is tempting to associate a single scale 
(=$\Lambda_{LFV}$) with all LFV operators, one should not use this as the
only criteria to judge the relative importance of some operator.
The possible values of the coefficient 
$\alpha^{ij}_n$ should also be taken into account to estimate
the strenght of the corresponding operator. Moreover, it could be the case
that the new physics associated to the scale $\Lambda_{LFV}$, could induce
operators with very different coefficients $\alpha^{ij}_n$.
This is already exemplified in the SM, where the GIM mechanism
allows the FCNC decay $b \to s+\gamma$ to be observable, but it
strongly suppresses the top decay $t \to c +\gamma$ \cite{ourbest}.

We are interested here in studying operators
that can induce LFV interactions of the Higgs boson.
In general, some of these operators will also induce LFV vertices
for the Z-boson ($Z l^+_i l^-_j$). 
Present data on LFV transitions can be used to bound their coefficients, 
mainly from the decays $l_i \to l_j +\gamma,
l_i \to l_j l_k l_k$, $Z\to l^+_i l^-_j$, $M \to l^+_i l^-_j$ 
(where $l_i= e, \mu, \tau$ and $M$ denotes the light mesons) 
and electron-muon conversion in nuclei.

Follwing Ref. \cite{areinwud}, one can specify whether some
operator arises as a tree-level or as a loop-effect from the
fundamental theory, and include for the last case the typical
loop-factor $1/16\pi^2$ to estimate its effect.
 At dimension-six there are a number of operators that can mix
different lepton flavours,
which can be classified  according to the type of bosonic field that is involved 
(gauge bosons and Higgs), as follows.

\begin{enumerate}

\item Operators that generate LFV photon interactions. These 
operators appear in two types:
\begin{eqnarray}
\label{eq2}
O^{ij}_{LW}&=&i(\bar{L}_i\tau^a\gamma_{\mu}D_{\nu}L_j)W^{a\mu\nu} \nonumber \\
O^{ij}_{LB}&=&i(\bar{L}_i\gamma_{\mu}D_{\nu}L_j)B^{\mu\nu} \nonumber \\
O^{ij}_{lB}&=&i(\bar{l}_{Ri}\gamma_{\mu}D_{\nu}l_{Rj})B^{\mu\nu}
\end{eqnarray}
and
\begin{eqnarray}
\label{eq3}
O^{ij}_{lW\phi}&=&(\bar{L_i}\sigma^{\mu\nu}\tau^al_{Rj})\Phi W^a_{\mu\nu} \nonumber \\ 
O^{ij}_{lB\phi}&=&(\bar{L_i}\sigma^{\mu\nu}l_{Rj})\Phi B_{\mu\nu}
\end{eqnarray}

\item LFV four-fermion operators, 
\begin{eqnarray}
\label{eq4}
O^{(1)}_{ijkl}&=&\frac{1}{2}(\bar{L_i}\gamma_{\mu}L_j)(\bar{L_k}\gamma^{\mu}L_l) 
\nonumber \\ 
O^{(3)}_{ijkl}&=&\frac{1}{2}(\bar{L_i}\gamma_{\mu}\tau^aL_j)(\bar{L_k}\gamma^{\mu}
\tau^aL_l) 
\end{eqnarray}

\item Operators that only generate LFV interaction of $Z$ and $H$,
\begin{eqnarray}
\label{eq5}
O^{ij}_{D\phi} &=& i(\phi^{\dagger}D^{\mu} \phi)(\bar{l}_{Ri}\gamma_{\mu} l_{Rj}) 
\nonumber \\
O^{(1)ij}_{D\phi} &=& i(\phi^{\dagger} D^{\mu} \phi)(\bar{L}_i\gamma_{\mu} L_j ) 
\nonumber \\
O^{(3)ij}_{D\phi} &=& i(\phi^{\dagger}\tau^a D^{\mu} \phi)(\bar{L}_i \gamma_{\mu}
 \tau^a L_j) ,
\end{eqnarray}
\begin{eqnarray}
\label{eq6}
O^{ij}_{Dl} &=& (\bar{L_i}D^{\mu}l_{Rj})D_{\mu}\Phi \nonumber \\
O^{ij}_{DL} &=& (\overline{D^{\mu}L_i}l_{Rj})D_{\mu}\Phi 
\end{eqnarray}

\item Yukawa-type operators that only generate LFV interactions of the Higgs, 
\begin{equation}
\label{eq7}
O^{ij}_{L\phi}= ( \Phi^{\dagger} \Phi ) (\bar{L_i} l_{Rj} \Phi) 
\end{equation}

\end{enumerate}

\bigskip

Several comments regarding these operators are in order

\begin{itemize}

\item The operators (2) and (3) must be generated as a loop-effect
in the fundamental theory. In this case
the constraints obtained from $e-\mu$ transitions
are quite strong, because they can induce LFV transitions 
of the photon, which mediates the decays $\mu \to e \gamma$ and $\mu \to e e e$. 
If one takes $\alpha^{ij}=1$, these processes implay an strong bound on
the scale associated to LFV ($\Lambda_{LFV} > O(100)$ TeV). 
However, the bounds associated with LFV transitions of the tau lepton are
significantly smaller.

\item The set (4) is favored in composite models, and it also implays strong bounds
on the scale  $\Lambda_{LFV}$ for $e\mu$ transitions.

\item The set of operators (5) and (6) 
provides a convenient framework to  discuss LFV for the Higgs sector, 
because one can use
the bounds on the decays $Z\to l_i l_j$ and $l_i \to l_j l_k l_k$
to constrain  their coefficients, and use them to predict the rates for the
Higgs decays $H\to l_i l_j$. 
Operators (\ref{eq5}) arise at tree-level, whereas (\ref{eq6}) arise as
one-loop effects from the underlying theory. It is interesting to note
that the operators (\ref{eq5},\ref{eq6},\ref{eq7}), which generate the vertices
$Zl^+_i l^-_j$ and $Hl^+_i l^-_j$, do not
induce the LFV photon vertex $A l^+_i l^-_j$, which in fact just follows
from $U(1)_{em}$ gauge invariance.

\item Operators (7) contribute to the fermion masses, and
in order to derive the corresponding Feynman rules in a
consistent manner, one should include their contribution
to the fermion mass matrices, then perform the diagonalization,
and finally write down the interaction lagrangian in terms of mass-eigenstates.
 However, we find
that the effect of the diagonalizaing matrices
can be re-absorved into the (unknown) coefficients of the
effective operators.  
\footnote{We have also verified, through a systematic use of the
equations of motion, that other operators of the type
$O_{ij}= i ( \Phi^{\dagger} \Phi ) \bar{L_i} \gamma^\mu D_{\mu} L_j$,
which appearently can induce LFV interactions for photon, Z and
Higgs boson, can be reduced to the form of eq. (7).}

\end{itemize}

 Since the operators (2-4) do not generate LFV
Higgs interactions, they willl not be discussed furthermore here.
Thus, our analysis will concentrate on the operators (5-7) 
\footnote{A complete analysis of LFV bounds, including a discussion of all the
one- and two-boson LFV vertices generated from these operators, 
will appear elsewhere \cite{ourfuture}.}. 

After SSB, one gets the LFV interactions of the Z and the remmainig Higgs
boson (H), which are described by the following lagrangians:
\begin{eqnarray}
\label{eq9}
{\it L}_{Zl_il_j}&=&-\frac{g}{4c_W}Z^{\mu}\lbrace \epsilon^{ij}_{D\phi}
(\bar{l}_{Ri}\gamma_{\mu}l_{Rj})+(\epsilon^{(1)ij}_{D\phi}+\epsilon^{(3)ij}_{D\phi})
(\bar{l}_{Li}\gamma_{\mu}l_{Lj}) \nonumber \\
&-& \frac{1}{(4\pi)^2}\frac{2\sqrt{2}c_W}{gv}[\epsilon^{ij}_{Dl}
(\bar{l}_{Li}\partial_{\mu}l_{Rj})+\epsilon^{ij}_{DL}(\partial_{\mu}\bar{l}_{Li}l_{Rj})]\rbrace+H.C.,
\end{eqnarray}
\begin{eqnarray}
\label{eq8}
{\it L}_{H^0l_il_j}&=&-g^2\epsilon^{ij}_{L\phi}H^0(\bar{l}_{Li}l_{Rj}) \nonumber \\
&+& \frac{i}{2v}\partial^{\mu}H^0[\epsilon^{ij}_{D\phi}(\bar{l}_{Ri}\gamma_{\mu}l_{Rj})
+(\epsilon^{(1)ij}_{D\phi}+\epsilon^{(3)ij}_{D\phi})(\bar{l}_{Li}\gamma_{\mu}l_{Lj})]
 \nonumber \\
&+& \frac{1}{(4\pi)^2}\frac{1}{\sqrt{2}v^2}\partial^{\mu}H^0
[\epsilon^{ij}_{Dl}(\bar{l}_{Li}\partial_{\mu}l_{Rj})+
\epsilon^{ij}_{DL}(\partial_{\mu}\bar{l}_{Li}l_{Rj})]+H.C.,                   
\end{eqnarray}
where the subindices $L,R$ are used to denote left- and right-handed fermion fields,
$v=\sqrt{2}<\Phi>_0$,  $c_W=cos\theta_W$, $s_W=sin\theta_W $. Also, in all the above 
operators, $L$ denotes the standard left-hand doublet of $SU(2)$.
In eqs. (8,9),  we have introduced the dimensionless quantities
$\epsilon^{ij}_n=(v/\Lambda_{LFV})^2\alpha^{ij}_n$, to absorb the dependence on the
parameters $\Lambda$ and $\alpha^{ij}_n$ into a single factor.
We have also introduced explicitly the factor $1/(4\pi)^2$,
for the coeffcients of the loop-induced operators
(\ref{eq6}). 
In fact, eq. (\ref{eq8}) describes the most general
LFV interactions of a Higgs boson that includes terms of 
dimensions four, five and six.
\footnote{In general, the parameters of the SM (dimension-four operators)
are modified by the dimension-six operators 
\cite{allops}. However, since there are no LFV effects at tree-level
in the SM, the contributions from the LFV operators considered here,
are of second order in the $\epsilon_n$ parameters and thus can be neglected.}

From the interactions contained in eqs. (8) and (9), we can derive the
expressions  for the decay widths of the 
processes $l_i \to l_j + l_k + l_k$ and $ Z \to l^+_i +l^-_j$, which
will be used to constrain the parameters $\epsilon^{ij}_n$.
We shall be working to first order in the parameters $\epsilon^{ij}_n$.
On the other hand, although there are strong experimental bounds on the 
radiative decays  $\mu \to e+\gamma, \tau \to e+\gamma, \mu+\gamma$,  
it happens that they receive contributions from the  operators (5-7) only
to second order in the parameters $\epsilon_n$ , thus the
resulting constraints are rather weak and can be neglected in the
present analysis. The results for the decay widths are:
\begin{eqnarray}
\label{eq10}
\Gamma (Z \to l^+_il^-_j)&=&\frac{\alpha m_Z}{96s_W^2c_W^2} 
 [|\epsilon^{ij}_{D\phi}|^2+|\epsilon^{(1)ij}_{D\phi}+\epsilon^{(3)ij}_{D\phi}|^2 
\nonumber \\
 & & +\frac{1}{4(4\pi)^4}|\epsilon^{ij}_{DL}- \epsilon^{ij}_{Dl} |^2],\\
\Gamma  (l_i \to l_j l^+_kl^-_k) &=& \frac{\alpha^2(g^{k2}_V+g^{k2}_A)m_i}
{2048\pi s_W^4c_W^4} (\frac{m_i}{m_Z})^4 \nonumber \\
& & \lbrace \frac{2}{3}[|\epsilon^{ij}_{D\phi}|^2+|\epsilon^{(1)ij}_{D\phi}+
\epsilon^{(3)ij}_{D\phi}|^2]+
\frac{1}{(4\pi)^4}\frac{m_i^2}{5m_Z^2}|\epsilon^{ij}_{Dl}+\epsilon^{ij}_{DL}|^2 
\nonumber \\
& & +\frac{1}{(4\pi)^2}\frac{\sqrt{2}}{3}\frac{m_i}{m_Z}Im\epsilon^{ij*}_{D\phi}
(\epsilon^{ij}_{Dl}+\epsilon^{ij}_{DL}) \rbrace, 
\end{eqnarray}
where $g^k_{V},g^k_{A}$ denote the vector and axial-vector couplings
of lepton $k$ with the Z boson; $\alpha$ denotes the fine structure constant,
$\theta_W$ is the electro-weak (EW) mixing angle. 
 In deriving these expressions,
we have neglected the lepton masses of the final states. For the decay
$l_i \to l_jl_kl_k$, we have included only the Z-mediated contributions, 
because we are assuming that the diagonal Higgs-fermion terms are of similar 
strength to the SM values, thus their contribution to the
decay amplitude can be neglected.
One can see inmediatly from the above expressions, that if all the operators
are present, then the effects coming from
the operator (\ref{eq5}) will be dominant, whereas the one coming from
the loop-induced operator (\ref{eq6}) will be very suppressed.

Then, we can also evaluate the width for the LFV Higgs decays
from the lagrangian (\ref{eq8}),
the resulting decay widths are given by:

\begin{equation}
\label{eq13}
\Gamma (H^0 \to l^+_i+l^-_j)  = \frac{\pi \alpha^2m_{H^0}}{s^4_W}
           [ |A^{ij}_L|^2+ |A^{ij}_R|^2 ]
\end{equation}

where

\begin{eqnarray}
\label{eq14}
A^{ij}_L  &=& \epsilon^{ij}_{L\phi} -
       \frac{s_W}{8c_W\sqrt{\pi\alpha}}
       [ \frac{m_i}{m_Z}(\epsilon^{(1)ij}_{D\phi}+\epsilon^{(3)ij}_{D\phi})
       -  \frac{m_j}{m_Z} \epsilon^{ij}_{D\phi} ] \nonumber \\
     & & -\frac{\sqrt{2}}{c^2_W} \frac{1}{(4\pi)^2} (\frac{m_{H^0}}{m_Z})^2
  [\epsilon^{ij}_{DL}+\epsilon^{ij}_{Dl} ] \\
A^{ij}_R  &=& \frac{s_W}{8c_W \sqrt{\pi\alpha} }
       [ \frac{m_i}{m_Z} \epsilon^{ij}_{D\phi}
       -\frac{m_j}{m_Z}(\epsilon^{(1)ij}_{D\phi}+\epsilon^{(3)ij}_{D\phi}) ] 
\end{eqnarray}

In order to present results for the branching ratios of the Higgs into the
LFV modes, we could choose many combinations for the $\epsilon$ parameters
to satisfy the experimental bounds on LFV proceses. However, in order to 
simplify the analysis, we shall only consider three scenarios, that we think 
illustrate the possibilities to detect LFV through Higgs decays, namely:

\begin{enumerate}

\item "Democratic" scenario, where we shall simply assume that 
all the parameters $\epsilon$ that appear in (10-11) are equal,
i.e. $\epsilon^{ij}_{D\phi}=\epsilon^{(1)ij}_{D\phi}=\epsilon^{(3)ij}_{D\phi}
=\epsilon^{ij}_{DL}= \epsilon^{ij}_{Dl}=\epsilon^{ij}$, 
which is probably the most conservative case. 
The resulting bounds arising from the $ Z \to l^+_il^-_j$ and
$l_i \to l_jl_kl_k$, are shown in  the first column of Table 1. 
We can see that the 3-body decay $\mu \to eee$ provides the strongest
bound for $e-\mu$ transitions, whereas the
Z-decays gives slightly better bounds for tau transitions. 
To evaluate the B.R. of the Higgs
boson in this "democratic" scenario we also take 
$\epsilon^{ij}_{L\phi}=\epsilon^{ij}$; the results are shown in Fig. 1,
and we can see that the mode $H\to \tau\mu/ \tau e$ reaches a B.R. of order
0.1 that seems to be at the reach of Tevatron for intermediate Higgs masses
(95 GeV $< m_H < 2m_W$ ),  whereas the decay $H\to \mu e$ 
can be at most of order $10^{-9}$.

\item "Loop-dominated" scenario, here we assume that only the 
loop-induced operators (6) contribute to LFV. In this case we find that
present data does not imposse strong bounds on the coefficients of these
operators, as it is shown in Table 1 (column 2). However, even if we assume
that the respective coefficients $\alpha^{ij}_n$ are of order 1, it is found that the
LFV Higgs decays are more suppressed: the B.R. turns out to be of order $10^{-3}$.
Notice that in this case we can not distinguish among the different LFV
modes.
 Fig. 2 shows the values of the B.R. for the LFV modes assuming several values
of the parameter $\epsilon^{ij}_{DL,Dl}$, (= $0.06, 0.005, 0.0005$, the first value 
corresponds to taking $\Lambda_{LFV}=1$ TeV and $\alpha^{ij}_{DL,Dl}=1$).

\item "Yukawa-dominated" scenario, in this case we assume that the operator (7)
is the only operator responsible for LFV Higgs interactions. This scenario
is very weakly constrained by low-energy bounds on LFV reactions, and in fact
the resulting bounds are not significant. Thus, we can evaluate
the B.R. for the Higgs LFV modes just as a function of the parameter 
$\epsilon^{ij}_{L\phi}$.
However, in order to introduce a criteria to discriminate among the
different decay modes, we shall also include a factor that takes into
account the breaking of the flavour symmetries that the operator
(7) would respect in the absence of fermion masses. The flavour-dependent
factor that we will use is similar to the one introduced in the general
Two-Higgs doublet model to suppress FCNC, which will be discussed in the
next section, namely we take:
$\epsilon^{23}_n=0.06 (\frac{m_\mu}{m_\tau})^{1/2}$,
$\epsilon^{13}_n=0.06 (\frac{m_e}{m_\tau})^{1/2}$,
$\epsilon^{12}_n=0.06 \frac{ (m_e m_\mu)^{1/2}}{m_\tau}$.
 The factor $0.06$ corresponds to the value of $\epsilon^{ij}_{L\Phi}$ 
that is obtained  with $\Lambda_{LFV}=1$ TeV and $\alpha^{ij}_{L\phi}=1$.
  The Results for the B.R. are shown in Fig. 3.
In this case the modes $H\to \tau \mu $ can also
reach a B.R. of order 0.1, which can be easily tested at 
the Run-II of Tevatron for intermediate Higgs masses.  

\end{enumerate}

\section{Results for specific models}

In this section we shall evaluate the LFV Higgs decays for
three specific models, namely for
the SM with massive neutrinos, the THDM-III and the
MSSM.

{\bf {a) LFV with massive neutrinos.}}
The SM with massive neutrinos provides the simplest
example where LFV decays of the Higgs boson appear.
It is well known that the existence of massive neutrinos requires
the inclusion of physics beyond the SM \cite{ramond,pilafts}. 
If we assume that the low-energy
effect of the physics responsible for neutrino masses is only
the diagonal neutrino mass matrix ($m_{li}$) and a leptonic mixing matrix
($V_{ij}$), then the leading contributions of massive neutrinos to the
 1-loop decay $H\to l^+_i l^-_j$ can be estimated by just
evaluating the loop with two internal W's, which gives the
finite result:
\begin{equation}
\label{eq16}
 \Gamma (H\to l^+_i l^-_j)=  \frac{\alpha^3  } {32\pi^3 \sin^6 \theta_W}
               \frac{m^2_i m^2_W}{m^3_H} |A_{ij}|^2 
\end{equation}

$m_i,m_W,m_H$ denote
the mass of the heavier lepton, W and Higgs boson, respectively. The factor
$A_{ij}$ is given by:
$ A_{ij}= V^{*}_{ik} V_{kj} \log \frac{m_{\nu k}}{m_W}$.

Present bounds on the mixing angles and the neutrino
masses \cite{kamiok}, can be satisfied with $A_{\tau\mu} \simeq O(1)$ \cite{pham}. 
Thus, for $m_H=100$ GeV 
the resulting B.R. for the mode $H\to \mu \tau$ is
 $B.R. \simeq 6 \times 10^{-7}$; whereas the mode
$H\to e \mu $ is expected to be much more suppressed.
These rates are clearly beyond experimental reach.

{\bf {b) LFV in the general Two-Higgs doublet model. }}
We are also interested in studying the 
LFV Higgs decays in the general two-Higgs doublet extension of the
SM (THDM). This model has potential problems with FCNC, which
were solved in its early versions (the so-called
Models I and II) \cite{weingla} by requiring
a discrete symmetry that restricted each fermion to couple
at most to one Higgs doublet. Then, Flavour changing 
neutral scalar interactions (FCNSI) are absent at the tree-level.
Later on, it was found that FCNSI could be
suppressed at acceptable rates, with relatively light Higgs bosons,
by impossing a more realistic pattern on the Yukawa matrices,
which in principle can be associated with some family symmetry
\cite{revmodIII}.

The phenomenological predictions of this model (called model III
in the literature \cite{oldmodIII})
have been studied to some extent. In this paper we shall use
the Higgs mass-eigenstate basis, which is more appropriate when one
is interested in the detection of direct Higgs signatures.
In this basis the LFV interactions of the neutral
Higgs boson $H^0$ of model III take the form:
\begin{equation}
\label{eq17}
{\cal{L}}_{LFV} = \xi_{ij} \cos \alpha \bar{l_i} l_j H^0 + h.c.
\end{equation}
where $\alpha$ denotes the mixing angle of the neutral Higgs sector, and 
$\xi_{ij}$ denotes the Yukawa coupling of the second
doublet.  
In order to satisfy the low energy data on FCNC, it was proposed
\cite{chengsher} the following ansazt :
\begin{equation}
\label{eq18}
 \xi_{ij}= \lambda_{ij} \frac{(m_i m_j)^{1/2}}{v}            
\end{equation}
where $v=246$ GeV and the lepton mass factor gives the order
of magnitude of the interaction. The coefficients
$\lambda_{ij}$ are dimensionless parameters that
can be constrained by comparing with present experimental
bounds on FCNC and LFV transitions.
The strongest bound for the parameters $\lambda_{ij}$ are
obtained from muon anomalous magnetic moment
\cite{sheretal},
namely: $  \lambda_{\mu \tau}  < 10$, which involves
only one coupling. 
 Other interesting bounds are:
 $( \lambda_{e\mu}\lambda_{\mu \tau} )^{1/2} < 5$, which 
is obtained from the decay $\mu \to e +\gamma$.
We have also studied the decay $\mu \to eee$, and find the result:
$(\lambda_{ee}\lambda_{e\mu} ) < 200$, which is not as good
as the one obtained from the muon anomalous
magnetic moment.

Then, the corresponding decay width for $h^0 \to l^+_i l^-_j$ is given by:
\begin{equation}
\label{eq19}
 \Gamma ( H^0 \to l^+_i l^-_j)=  \frac{\xi^2_{ij} }{8\pi} \cos^2 \alpha m_H
\end{equation}
The dominant decay mode of the Higgs boson
in the intermediate mass range is
expected to be $H \to b \bar b$, which will be proportional to
$\sin^2 \alpha$.

Then, if we include the present bounds for $\lambda_{ij}$
, the resulting upper limit on the B.R. of the mode $H\to e \mu $
is of the order $10^{-5}$ for $\sin\alpha=0.1$, which does not
seem to be at the reach of future experiments. 
On the other hand, the mode
 $H\to \mu \tau$ is allowed to have a  B.R. of order  0.1, which seems
at the reach of future colliders. In fact, the direct search
for this mode at Run-II of Tevatron could be used to
improve the limits on the values of $\lambda_{\mu\tau}$.
 Results for $\sin\alpha=0.1$ (0.9) are shown in table 3.

{\bf {c) The Minimal SUSY SM (MSSM).}}
LFV interaction can also arise at 1-loop in the
minimal SUSY standard model (MSSM) \cite{susyrev}. 
The MSSM has become one the most preferred extension of the 
SM; it predicts new signatures associated with
the superpartners, although it reproduces the SM agreement with
data.
The most general lagrangian for the MSSM has problems with FCNC.
In the supergravity (SUGRA) inspired models \cite{sugrarev},
potential FCNC problems are solved by assuming
that the sfermion masses are degenerated and the
cubic A-terms are proportional to the mass of the corresponding fermion.
However, these conditions only hold at a heavy (GUT) scale, and
once these parameters are evolved down to the EW scale ($M_{EW}=246$ GeV),
some detectable FCNC effects may arise. 
To illustrate the rates that result in the MSSM, we shall study
the LFV effects that arise from the Higgs-slepton vertex,
using the results of \cite{japanLFV}, which finds the following
expressions for the cubic A-terms in the SUSY $SU(5)$ model,
\begin{equation}
\label{eq20}
 A^l_{ij} =  -\frac{9}{16\pi^2 }
  [ V^{3i}_{CKM} V^{3j*}_{CKM} Y^2_t Y_i ]a_0 m_0 
\log \frac{M_{EW}}{ M_{GUT}}
\end{equation}
where $V^{ij}_{CKM}$ denotes the CKM mixing matrix, 
whereas $Y_f,a_0, m_0$ correspond to the Yukawa, trilinear and universal
scalar mass, respectively.

To estimate the decay width $H^0 \to l^+_i l^-_j$
for the light Higgs boson $H^0$, one could evaluate 
the graph that includes only sleptons-bino inside the loop, 
which gives the following finite result:
\begin{equation}
\Gamma(H^0 \to l^+_i l^-_j)= \frac{9\alpha^3}{6400 \pi^2 c^2_W}
                              [\frac{ A^{l2}_{ij} m^3_H}{ {\tilde{m_0}}^3}]
\end{equation}
Then, if we include the corresponding 
lepton masses and take $a_o=\tilde{m_0}= 100$ GeV, the resulting  B.R.
for $H\to e\mu$ is of the order $10^{-11}$, whereas for $H\to \tau \mu/ \tau e$
it can be of order $10^{-6}$, which does not seem to be at the reach of
future colliders.

\section {Discussion of results and conclusions.}

We have studied the Lepton Flavor Violating (LFV) decays of Higgs bosons
$H\to l^+_i l^-_j$, in several extensions of the SM. Results for
the Effective lagrangian approach and the general two-Higgs doublet
model (THDM-III)
are presented in detail, whereas for the SM with massive neutrinos
and the Minimal SUSY-SM,
some estimates are given.
In the effective lagrangian case, it is found that
for those operators whose coefficients can not be constrained by
present data, the LFV higgs decays could have
a Branching Ratio (B.R.) of order $10^{-1}$, which could be detected at 
hadron colliders (Tevatron or LHC). Those cases when
current bounds on LFV transition apply, induce
strong bounds for $e-\mu$ transitions, and the resulting B.R. for
the decay $H\to e\mu$ is of order $10^{-9}$; however
the bounds involving the tau are considerably weaker,
and allow the mode 
$H\to \tau \mu/\tau e$ to have a B.R. of the order $10^{-1}$.
Similar results are obtained for THDM-III; whereas the
corresponding decays in the SM with massive neutrinos and
the minimal SUSY-SM are much smaller, and beyond
experimental reach.

Thus, we found that the LFV Higgs decays $H\to \tau \mu/\tau e$ can have large
branching ratios, of order 0.1 in some cases. It is important to estimate
weather such rates can be detected at future colliders. At the next Run-II
of Tevatron, it is possible to use the gluon-fusion mechanism to produce a
single Higgs boson; assuming that the production cross-section is of
similar strenght to the SM case, about 1.2 pb for $m_H=100$ GeV, it will
allow to produce 12,000 Higgs bosons with an integrated luminosity of 10
fb-1. Thus, for $B.R.(H\to \tau \mu/\tau e) \simeq  \times 10^{-1}$
 Tevatron can produce 1200 events.
Since it has been shown possible to identify the hadronic decays of the
tau lepton at Tevatron \cite{hpmlim}, it seems factible to study the 
LFV Higgs into $\tau \mu / \tau e$ by impossing a cut on the 
invariant-mass of the final state, which
should allow to identify the signal. On the other hand, since the 
mode $H \to \mu e$ can reach at most a B.R. of
order $10^{-3}$, then it will require a higher luminosity to be detected.
For instance, if we have an integrated luminosity of $30 fb^{-1}$, then we
get 18 events, after including a detection efficiency of $50 \%$. Although the
rate will be low, the signal is so distinctive that using a cut on the
invariant mass should allow to eliminate the SM backgrounds.

In summary, we have found several cases where the LFV higgs decays
may be at the reach of the Run II of Tevatron, which should be
considered as a motivation, or starting point,
to search for LFV Higgs decays at present and future colliders.

{{\bf Acknowledgment.-} Valuable discussions with G. Kane, 
H.J. He and C.P. Yuan are acknowledged. JLDC acknowledges 
the kind hospitality of the theory group of LBL (Berkeley) 
and the finantial support of {\it Fundaci\' on M\' exico-USA 
para la Ciencia}. This work was supported by CONACYT and 
SNI (M\'exico).}

\newpage

\begin{figure}[!hbt]
\begin{center}
\epsfysize=8cm \epsfbox{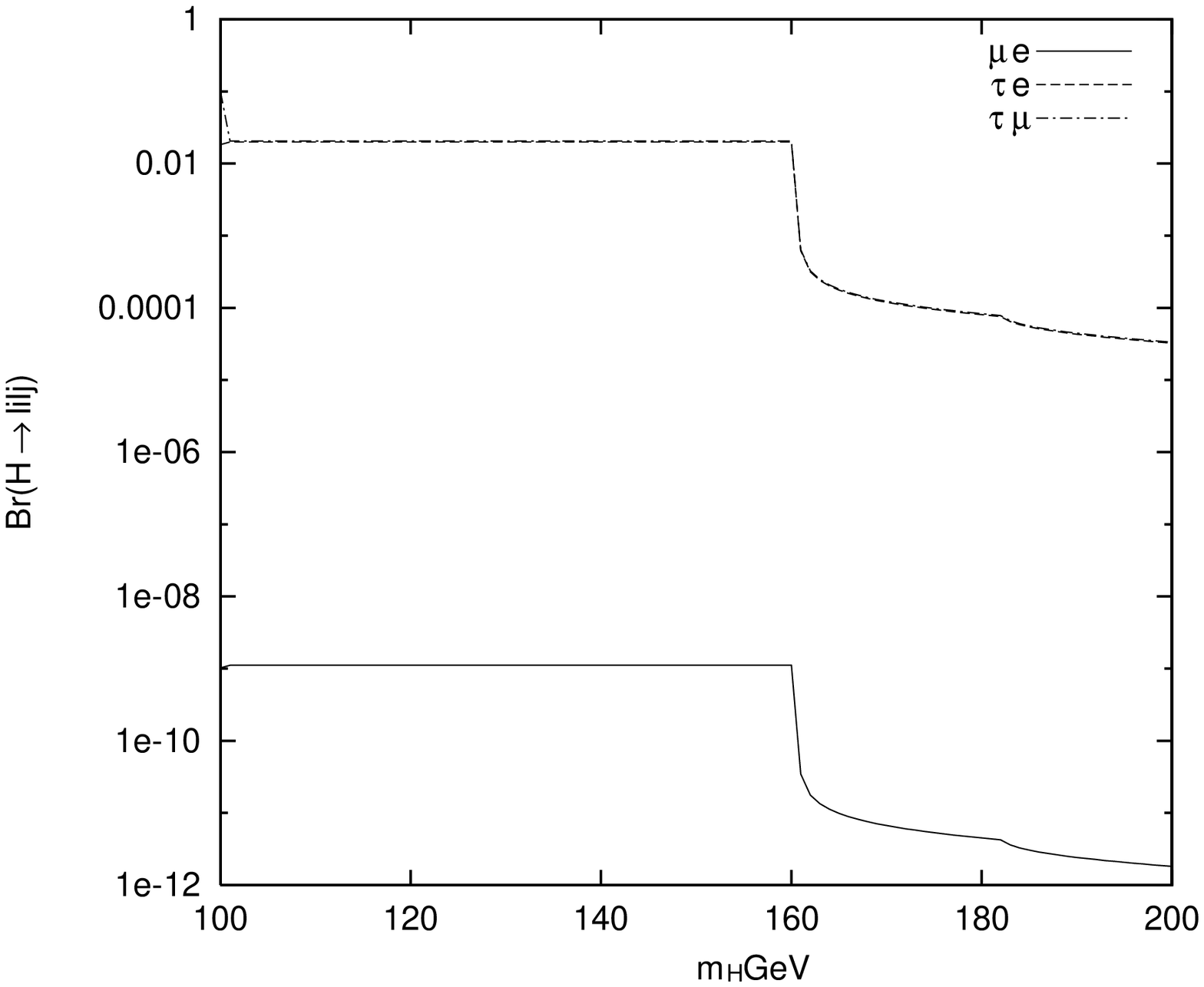}
\end{center}
\caption{B.R. of LFV Higgs decays for the effective lagrangian
approach in the "democratic scenario". The splitting between the $\tau \mu$ and
$\tau e$ curves is about $10^{-4}$.}
\end{figure}

\begin{figure}[!hbt]
\begin{center}
\epsfysize=8cm \epsfbox{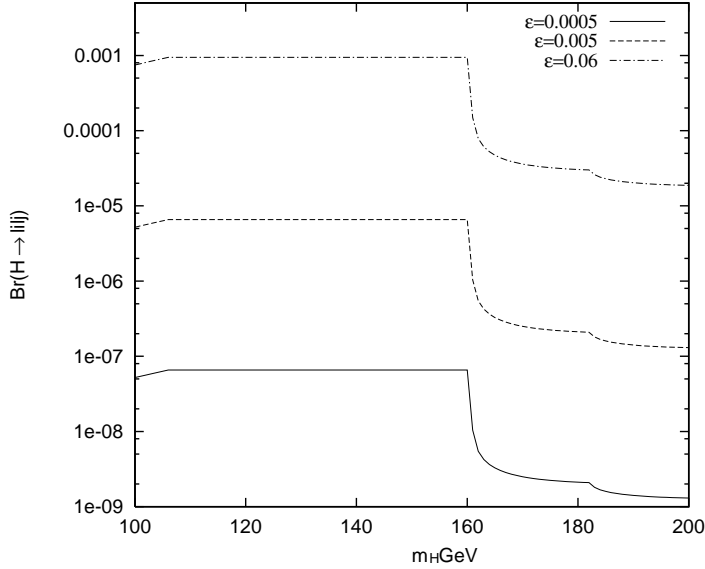}
\end{center}
\caption{ B.R. of LFV Higgs decays for the effective lagrangian approach
in a "Radiative-dominated" scenario. In this case all the LFV Higgs decay
modes have the same B.R..}
\end{figure}

\begin{figure}[!ht]
\begin{center}
\epsfysize=8cm \epsfbox{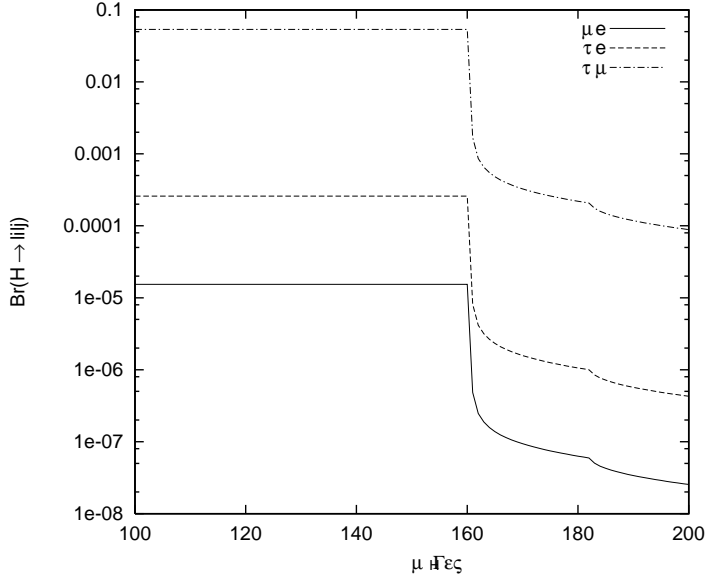}
\end{center}
\caption{ B.R. of LFV Higgs decays for the effective lagrangian approach
in a "Yukawa-dominated" scenario.}
\end{figure}

\vspace{10cm}

{\bf TABLE CAPTION}

\bigskip

Table 1. Bounds on the LFV parameters $\epsilon^{ij}_n$ for the effective
lagragian scenarios considered in this paper.

\bigskip

Table 2. B.R. of LFV Higgs decays for the THDM-III.
Results are shown for $\sin\alpha=0.1$, and
the numbers in paranthesis correspond to $\sin\alpha=0.9$.


\bigskip

\begin{center}
\begin{tabular}{||l|l|l|l|}
\hline
    Mode   &   "Democratic" scenario     &  "Loop" scenario & 
"Yukawa" scenario \\
           & $|\epsilon^{ij}|^2<$ & $|\epsilon^{ij}_{Dl}-\epsilon^{ij}_{DL}|^2<$ & 
    \\ \hline 
 $Z \to \mu e$    &    $ 1.1 \times 10^{-5}$    &   5.4 &
  --- \\ \hline
 $Z \to \tau e$   &    $ 6.2 \times 10^{-5}$    &   31.2 &
  --- \\ \hline
$Z \to \tau\mu$   &    $ 1.1 \times 10^{-4}$    &   54.2 &
  --- \\  \hline
    Mode   &  $|\epsilon^{ij}|^2<$ & $|\epsilon^{ij}_{Dl}+\epsilon^{ij}_{DL}|^2<$  &
  $|\epsilon^{ij}_{L\Phi}|^2/ m^4_H<$  \\ \hline

 $\mu \to eee$    &   $4.4 \times 10^{-12}$      &  1.4   &
    $3.42 \times 10^{-9}$  \\  \hline
 $\tau \to e ee$   &   $8.1 \times 10^{-5}$      &  $9.3 \times 10^4$  &
     $6.3 \times 10^{-2}$  \\ \hline
 $\tau \to \mu ee$   &   $8.1 \times 10^{-5}$      &  $9.6 \times 10^4$  &
     $ 6.5 \times 10^{-2}$  \\  \hline
 $\tau \to \mu \mu \mu $   &   $8.1 \times 10^{-5}$      &  $5.4 \times 10^4$  &
 $8.5 \times 10^{-7}$  \\ \hline

\end{tabular}
\end{center}

\bigskip

{\bf \, \, \,  Table. 1}

\bigskip

\bigskip

\begin{center}
\begin{tabular}{||l|l|l|}
\hline
$m_H$ GeV & $B.R.(H\to \mu \tau )$  &  $B.R.(H\to e \mu)$ \\
\hline 
 100. &   0.7 (0.1)     &  $1.3\times 10^{-5}$ ($2.0\times 10^{-6}$) \\
\hline
 130  &  0.7 (0.1)      &  $1.2\times 10^{-5}$ ($2.1\times 10^{-6}$) \\
\hline
 170. &  0.3 ($1.2\times 10^{-3}$) & $5.5 \times 10^{-6}$ 
 ($2.3 \times 10^{-8}$) \\
\hline
 200. &  0.1 ($3.5 \times 10^{-4}$) & $2.2 \times 10^{-6}$ 
($6.4 \times 10^{-9}$) \\
\hline
\end{tabular}
\end{center}

\bigskip
{\bf  \, \,  \, Table. 2} 

\end{document}